\documentclass{article}
\usepackage[T1]{fontenc} % add special characters (e.g., umlaute)
\usepackage[utf8]{inputenc} % set utf-8 as default input encoding
\usepackage{ismir,amsmath,cite,url}
\usepackage{graphicx}
\usepackage{color}
\usepackage{booktabs}
\usepackage{placeins}
\usepackage{subcaption}
\title{Mel Spectrogram Inversion with Stable Pitch}

\threeauthors
  {Bruno Di Giorgi$^*$} {Apple \\ {\tt bdigiorgi@apple.com}}
  {Mark Levy$^*$} {Apple \\ {\tt mark\_levy@apple.com}}
  {Richard Sharp} {Apple \\ {\tt richard\_sharp@apple.com}}

\def\authorname{B. Di Giorgi, M. Levy, and R. Sharp}

\usepackage[bookmarks=false,pdfauthor={\authorname},pdfsubject={\papersubject},hidelinks]{hyperref}
\usepackage{breakurl}
\begin{document}

\maketitle
\def\thefootnote{*}\footnotetext{Equal contribution}\def\thefootnote{\arabic{footnote}}

\begin{abstract}

% context: neural vocoders
Vocoders are models capable of transforming a low-dimensional
spectral representation of an audio signal, typically the mel spectrogram, 
to a waveform. Modern speech generation pipelines use a vocoder 
as their final component.
Recent vocoder models developed for speech achieve a high degree of realism,
such that it is natural to wonder how they would perform on music signals.
% problem statement
Compared to speech, the heterogeneity and structure of the musical sound 
texture offers new challenges.
In this work we focus on one specific artifact that some vocoder models designed for speech
tend to exhibit when applied to music: 
the perceived instability of pitch when synthesizing sustained notes. 
We argue that the characteristic sound of this artifact is due to
the lack of horizontal phase coherence, which is often the result of 
using a time-domain target space with a model that is invariant 
to time-shifts, such as a convolutional neural network.

% contributions
We propose a new vocoder model that is specifically designed for music.
Key to improving the pitch stability is the choice of 
a shift-invariant target space that consists of the magnitude spectrum and the phase gradient.
We discuss the reasons that inspired us to re-formulate the vocoder task, 
outline a working example, and evaluate it on musical signals\footnote{
  Example reconstructions \burl{https://machinelearning.apple.com/research/mel-spectrogram}}.
Our method results in 60\% and 10\% improved reconstruction of 
sustained notes and chords with respect to existing models, 
using a novel harmonic error metric.

\end{abstract}
\section{Introduction}\label{sec:introduction}

In modern speech synthesis pipelines a first model generates a 
low-dimensional audio representation, usually the mel spectrogram,
from text; and a second model, named Vocoder, 
transforms the mel spectrogram to an audio waveform.
% Theoretically the same design can be applied to the music signal,
% however closer inspection reveals challenges that are exclusive to the
% music domain. In this work we focus on the Vocoder task of recovering
% the audio from the mel spectrogram.
Theoretically, vocoders designed for speech could be directly 
applied to musical signals; however closer inspection reveals 
features and constraints that are exclusive to the music domain.
% Recent vocoders for speech
For example, unlike speech, music signals can be polyphonic
and contain longer sustained notes whose pitch precision and stability
is essential.

The stability of a sustained pitched note manifests 
in the time-domain audio signal 
as the steady repetition of a periodic waveform.
Periodic patterns are by definition not shift-invariant, except 
for shifts of an integer number of periods, therefore they
require some form of auto-regression in order to be reproduced accurately.
As expected, time-domain vocoders using shift invariant architectures 
\cite{kumar2019melgan,kong2020hifi}, despite other advantages
such as generation efficiency, 
produce jitters 
that are perceived as pitch and timbre instability.
For this reason, other time-domain generative models for audio 
include an autoregressive mechanism in the neural architecture 
\cite{oord2016wavenet, kalchbrenner2018wavernn, morrison2021chunked}.
In practice, time-domain models are required to learn
all possible shifts of periodic patterns, 
a space that increases exponentially for polyphonic music, and 
how to create smooth sequences of these patterns.

\begin{figure}[t]
  \centering
  \includegraphics[width=0.95\columnwidth,trim={0 500pt 0 0},clip]{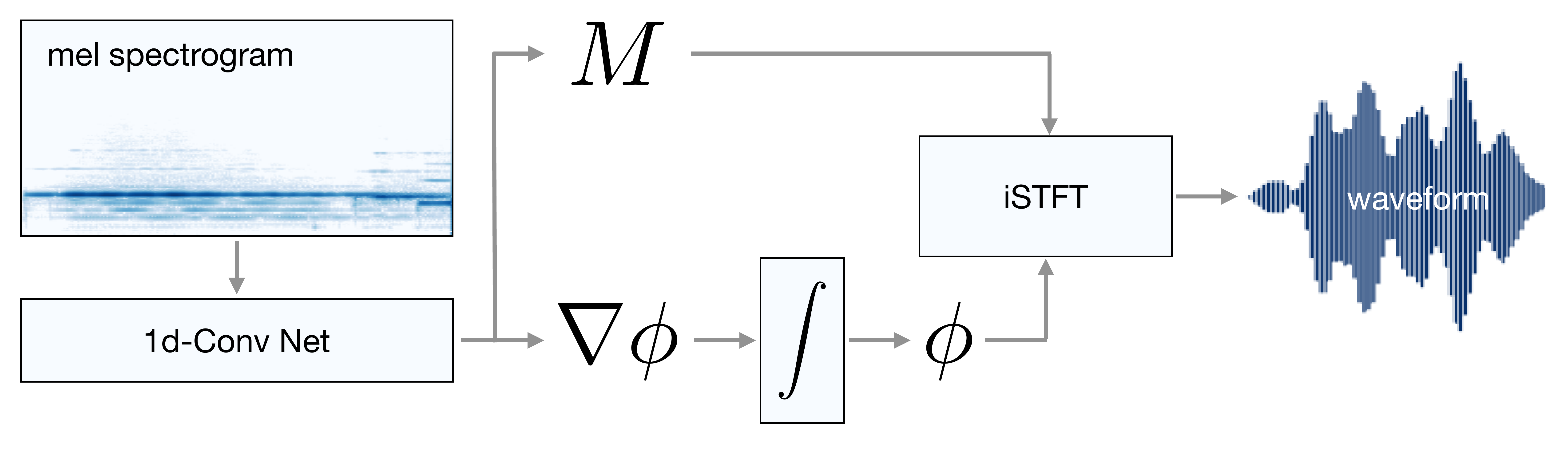}
  \caption{Our proposed model for mel spectrogram inversion. A one dimensional CNN estimates 
  the magnitude and the phase gradient from the mel spectrogram. The phase gradient 
  is then integrated to estimate the phase spectrum and finally audio is obtained via
  the inverse STFT.
  }
  \label{fig:system}
\end{figure}

Inspired by a recent generative model for single notes \cite{engel2019gansynth}, 
we propose a new vocoder model for music (Fig. \ref{fig:system}), where the target of the neural network is
an intermediate frequency-domain audio representation that is shift-invariant for sustained
notes. This representation is composed of the magnitude spectrum and the phase
gradient, and can be later turned to audio via: 
1.\ a phase integration algorithm and 2.\ the inverse STFT.
The proposed design can be used with an efficient shift-invariant neural architecture 
and still yield stable reconstruction of sustained notes. Specifically, our contributions 
include:
\begin{itemize}
  \setlength\itemsep{-0.1em}
  \item a formulation of the mel spectrogram inversion task, matching 
  shift-invariant network and target, in order to improve the perceived stability of sustained notes
  % TODO: The next is not evaluated and may be considered part of the re-formulation - remove?
  \item a phase integration algorithm  
  \item an evaluation metric measuring pitch stability for multiple notes
\end{itemize}

% Speech synthesis pipelines
% Generation to low-dim -> mel-spectrogram inversion

% How is music different from speech

% Literature on music models
% drum loops (hung2021benchmarking)

% Our contributions

% our contributions:
% 1. new output space using phase gradient
% 2. new phase integration algorithm
% 3. error metric that is able to show these errors

% Results hints
% summary
% The remainder of this paper is organized as follows: 
% Section \ref{sec:background} gives an overview of the techniques and the
% proposed formulation of the mel spectrogram inversion task;
% Section \ref{sec:model} describes our model; 
% we outline the experiments in Section \ref{sec:experiments}; finally,
% in Section \ref{sec:conclusions} we summarize our work and 
% propose directions for future research.

% Paper structure
% TODO: optional demo website

\section{Background}\label{sec:background}
% The DSP background needed for the paper

% i: sample index / summation variable
% n: frame index
% m: frequency index
% R: hopsize
% N: framesize and window length

% phase
% phi'_\omega: phase derivative along the frequency axis computed by the Auger-Flandrin technique
% phi'_t: phase derivative along the time axis computed by the Auger-Flandrin technique
%    note this is not the instantaneous angular frequency, but the IAF minus the bin's nominal angular frequency
% \Delta{n} = - phi'_\omega / R   time bin-offset
% \Delta{m} = phi'_\t * N / 2pi   freq bin-offset
% \frac{d\phi}{dm} = - \tilde{n} 2piR/N = phi'_\omega 2pi / N  phase shift along frequency
% \frac{d\phi}{dn} = (m + \tilde{m}) 2piR/N = (\omega_m + phi'_t) * R  phase shift along frequency

% simplification: update to assign symbols phi'_\omega and phi'_t to the simplest
% expression, obtainable also from first order differences
% check Equations.ipynb notebook for equation tests
% phi'_\omega: local group delay [samples]
% phi'_t: instantaneous angular frequency (phase shift per sample rad/sample)
% \Delta{n} = - phi'_\omega \frac{N} / {2 \pi R} [time-bins]  time bin-offset
% \Delta{m} = (phi'_\t - \omega_m) * N / 2pi  [freq-bins]  freq bin-offset

% Instantaneous angular frequency and local group delay
% cite FulopFitz2006AlgorithmsForComputingThe

% TODO: use "log-magnitude mel spectrogram (log-mel spectrogram)"

\begin{figure}[ht]
  \centering
  \includegraphics[width=\columnwidth]{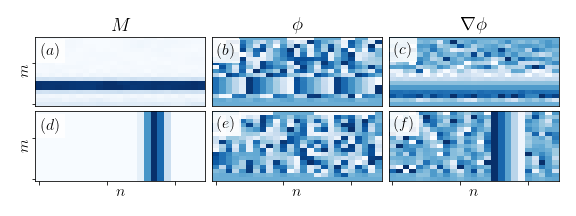}
  \vspace{-1em}
  \caption{
    Magnitude $M$, phase $\phi$ and phase gradient $\nabla\phi$ patterns for a
    sinusoidal (top row) and an impulse (bottom row)
    signal. While the magnitude spectrum is easy to interpret visually, 
    patterns in the phase spectrum are harder to decipher, but become
    evident in the phase gradient.
    }
  \label{fig:phase-patterns}
 \end{figure}

% Note: this formulation has centered phases, so it works with the
% equations for the local group delay
A discrete audio signal $x$ can be analyzed in the time-frequency space using the STFT:
\begin{equation}
X[m, n] = \sum_{i}x[i + nR]w[i]e^{-j\omega_m i},
\end{equation}
where $m$ and $n$ are the integer frequency bin and time frame indices, $R$ is the hop size 
between successive frames, $w$ is a window function defined in the 
$[-N / 2, N / 2)$ interval, with $N$ being the frame size and $\omega_m=2 \pi m / N$ the 
angular frequency.
The STFT is a complex-valued matrix, as such it can be represented in the polar form:
\begin{equation}
X[m, n] = M[m, n]e^{-j\phi[m, n]}.
\end{equation}

\noindent The magnitude component $M$ highlights the energy of the signal at various locations in the time-frequency grid:
it is easier to interpret and more widely used than the phase component $\phi$. 
However, the phase spectrum is of primary perceptual importance
to reconstruct the audio signal precisely, and while 
harder to interpret at first sight, it does contain patterns that can guide model design choices (Fig. \ref{fig:phase-patterns}).

In Sect. \ref{ssec:sinimp} we describe two 
patterns that form in the magnitude and phase spectrum corresponding to the occurrence of 
ideal sinusoidal and impulsive signal components.

\subsection{Sinusoidal and impulsive components}\label{ssec:sinimp}

\subsubsection{Sinusoidal components}\label{sssec:sin}

In the magnitude spectrum, sinusoidal components such as any single harmonic of a 
pitched instrument's sustained note, show up as horizontal lines (Fig. \ref{fig:phase-patterns}(a)).
While the magnitude spectrum does not depend on the frame index $n$,
and is therefore shift-invariant, the phase spectrum depends linearly on $n$ 
(Fig. \ref{fig:phase-patterns}(b)), and the rate of change is 
given by the frequency of the sinusoidal component.
Failing to reconstruct this linear relation between phase and time
results in loss of horizontal phase coherence, perceived as unstable pitch, 
because errors are attributed to sudden changes of the 
frequency of the sinusoidal component.

\subsubsection{Impulsive components}\label{sssec:imp}

Impulsive components such as the attack of a percussion instrument show up as 
vertical lines in the magnitude spectrum (Fig. \ref{fig:phase-patterns}(d)).
While the magnitude spectrum does not depend on the frequency index $m$,
the phase spectrum depends linearly on $m$ (Fig. \ref{fig:phase-patterns}(e)), and
the rate of change depends on the offset between the location of the impulse and the 
frame center.
Failing to reconstruct this linear relation between phase and frequency results in 
loss of vertical phase coherence, which is perceived as smeared transients, 
because the errors are attributed to the location of the impulse.

% % This paragraph could be expanded to a section of its own
% Many insights in the structure of the phase spectrum have been developed
% in the Time Scale Modification (TSM)
% literature.

% ---------------
\subsection{Phase gradient}\label{sec:phasegrad}

The linear patterns that emerge in the phase for sinusoidal and impulsive 
components are better highlighted in the two components of the phase gradient 
$\nabla\phi = (\phi'_i, \phi'_m)$.

The partial derivative of phase along the time dimension $\phi'_i$ is called \emph{instantaneous
frequency}. For the bins that belong to sinusoidal components, 
$\phi'_i$ is constant and the phase can be propagated horizontally:

\begin{equation}\label{eq:phigrad-if}
\phi[m, n + 1] = \phi[m, n] + R \phi'_i[m, n].
\end{equation}

% Instantaneous angular frequency \phi'_i  rad/sample
\noindent The partial derivative along the frequency dimension $\phi'_m$ is called \emph{local 
group delay}. For the bins that belong to impulsive components,
$\phi'_m$ is constant and the phase can be propagated vertically:

\begin{equation}\label{eq:phigrad-lgd}
\phi[m + 1, n] = \phi[m, n] + \phi'_m[m, n].
\end{equation}

\noindent In the time-frequency reassignment literature (see e.g.\ \cite{fitz2009unified}), 
the phase gradient components are used to assign the energy of a spectral bin $(m, n)$
to a nearby point of maximum contribution $(\dot{m}, \dot{n})$:
\begin{equation}
  \label{eq:n-bin-offset-dot}
  \begin{array}{r@{}l}
    \dot{m}[m, n] &= m + \Delta{m}[m, n]\\[8pt]
    \dot{n}[m, n] &= n + \Delta{n}[m, n],
  \end{array}
\end{equation}
\noindent where $\Delta{n}[m, n]$ (Fig. \ref{fig:phase-patterns}(f)) and $\Delta{m}[m, n]$ (Fig. \ref{fig:phase-patterns}(c)) represent
time and frequency bin offsets and are derived from the phase gradient
\begin{equation}
  \label{eq:n-bin-offset-deltas}
  \begin{array}{r@{}l}
    \Delta{m}[m, n] &= \phi'_i[m, n] \frac{N}{2\pi} - m\\[8pt]
    \Delta{n}[m, n] &= -\phi'_m[m, n] \frac{N }{2 \pi R}.
  \end{array}
\end{equation}

\noindent In the following sections, we discuss how the phase gradient can be used for
mel spectrogram inversion.

\subsection{Mel spectrogram inversion}\label{ssec:mel}

The log-amplitude mel spectrogram (simply mel spectrogram from now on) 
is a low-resolution time-frequency representation that is derived from the 
power spectrogram $M^2$, 
by first warping the frequency axis using the mel scale, then scaling the values to log-amplitude. 
Estimating the original audio signal $x$ from the mel spectrogram
requires recovering the information that has been lost in the direct computation, i.e.\
the phase information and the linearly-spaced and higher frequency resolution of the magnitude spectrum.

While the majority of the recent approaches try to learn this inverse transformation end-to-end, 
this is especially hard for a polyphonic music signal. 
To precisely reproduce a sustained note, an end-to-end model needs to learn: 1.\ different patterns
for every combination of phase shift and period of a periodic waveform, and 2.\ how to
activate them in the right sequence \cite{engel2019gansynth}. 
Accomplishing both tasks is challenging for speech and arguably even more so for music, 
which contains generally longer pitched sounds with wider pitch range, 
possibly multiple concurrent fundamental frequencies (polyphony), and 
whose absolute precision is essential.

Instead of reconstructing the signal in the time domain, we propose to use as output space an 
intermediate time-frequency representation consisting of three channels: the magnitude spectrum 
and the two components 
of the phase gradient: $(M, \phi'_i, \phi'_m)$.
The phase gradient is later integrated to estimate the phase spectrum $\phi$, and finally 
audio is computed via the inverse STFT.

A model trained on our proposed output representation does \emph{not} need to learn: 1.\ the
shift variations of periodic waveforms as those are explicitly modeled by the inverse STFT, and 2.\ 
how to sequence phase, which is handled via the phase integration algorithm.
Differently from the phase spectrum, the phase derivative along time is shift-invariant,
thus it is a more suitable target for a shift-invariant architecture, such as the 
convolutional neural network.

% similar ideas from the literature
The approaches that have been suggested in the recent years for neural audio synthesis
in the time-domain have to use auto-regression to achieve horizontal phase coherence.
For example, autoregression is at the core of models like 
WaveNet \cite{oord2016wavenet} and 
WaveRNN \cite{kalchbrenner2018wavernn},
however the fact that it is applied at the rate of audio samples make these model 
prohibitively expensive for the generation of high-resolution audio signals.

Audio domain shift-invariant convolutional neural vocoders can 
generate audio samples with much higher efficiency, 
but are not suited to reconstruct long pitched components precisely. 
This holds regardless of the training strategy, and includes for example
generative adversarial networks (GAN) based models \cite{kumar2019melgan, kong2020hifi} 
and diffusion based models \cite{kong2020diffwave, kandpal2022music}.
A recent neural vocoder for speech mel spectrogram inversion \cite{morrison2021chunked}
adds an autoregressive loop that works on chunks of audio. 
% rephrasing given review
The autoregressive nature of this architecture allows 
performing temporal integration, the operation needed to reconstruct 
stable sinusoidal components, while advancing by audio chunks rather than samples improves efficiency.
% This design allows it to learn 
% an implicit representation of phase and enhances the reconstruction of sinusoidal components,
% while maintaining higher generation efficiency compared to models that are auto-regressive 
% at the level of audio samples. 
However, the poor reconstruction quality observed when applying this model to music signals suggests that
it is difficult to learn signal properties such as the rotation of phase from data with sufficient generalization.

In the neural audio synthesis literature, using instantaneous frequency has been 
considered explicitly in \cite{engel2019gansynth}, where it is generated alongside the magnitude spectrum
in order to reconstruct the audio of single notes, conditioned on the pitch contour and a 
timbre embedding.

% First, it requires the network to implicitly learn some form of inverse STFT transform, effectively
% wasting capacity for a transformation that is available and relatively fast to compute.
% Second, in order to smoothly reconstruct the sinusoidal components, the network needs to maintain a state, 
% which requires some form of feedback loop, such as recurrent layers or an otherwise auto-regressive 
% architecture \cite{morrison2021chunked}.

\subsection{Phase integration}\label{ssec:phaseintegration}

% Time-stretching algorithms based on the Phase Vocoder \cite{dolson1986phase} (PV) use the 
% phase derivatives 
% of the input signal to produce a scaled version that maintains the same 
% perceptual qualities \cite{damskagg2017audio}, \cite{pruvsa2016real}. 

Recovering the phase spectrum from the phase gradients requires an integration step.
% Integrating a 2d function from the sampled gradients
Theoretically, perfect integration should be possible under specific constraints, 
such as continuous phase gradient spectrum and window functions with infinite support. 
In practice, however there is no
closed-form solution for integrating typical discrete phase gradient spectra \cite{pruvsa2016real}.

% Phase integration in the Phase-Vocoder TSM literature
Well-known phase gradient integration algorithms have been developed in the 
Time-Scale Modification (TSM) literature. The standard Phase-Vocoder (PV) algorithm propagates the phase derivative along the time 
dimension to 
modify the duration of sinusoidal components \cite{dolson1986phase}.
% problems of the standard vocoder
The PV is able to preserve horizontal phase coherence, but struggles with vertical phase
coherence, leading to smeared transients.
Improvements to the standard phase-vocoder algorithm \cite{pruuvsa2017phase, 
laroche1999improved, damskagg2017audio} 
use the magnitude spectrum to identify sinusoidal and/or impulsive components, and 
can propagate phase in either direction (time or frequency) depending on the local
properties of the signal.

% Ref to our proposed algorithm
The phase gradient integration algorithm that we develop (Sect. \ref{ssec:methodphase}) 
is inspired by
these recent variations, and leads to subjectively improved reconstruction
quality, alongside increased computational efficiency. 
A formal evaluation of the integration algorithm is out of the 
scope of this manuscript and left as future work.

% \subsection{Vocoders}\label{sec:related}

% % Car-GAN 
% \cite{morrison2021chunked}
% % GAN Synth
% \cite{engel2019gansynth}
% % Sinusoidal Model (is monophonic, no machine learning, poor evaluation)
% \cite{natsiou2021sinusoidal} 
% % ddsp
% \cite{engel2020ddsp}

\section{Model} \label{sec:model}

In this section we describe our proposed model for mel spectrogram inversion.
The model is composed of a time-wise convolutional neural network (Sect. \ref{ssec:network})
that estimates magnitude and phase gradient from the mel spectrogram, and 
a phase integration algorithm (Sect. \ref{ssec:methodphase}) that estimates the phase 
spectrum given the phase gradient.
The time-domain reconstructed audio is finally obtained via inverse STFT from the magnitude and 
phase spectra.

\subsection{Network architecture} \label{ssec:network}

\begin{figure}[t]
  \centering
  \includegraphics[width=0.7\columnwidth]{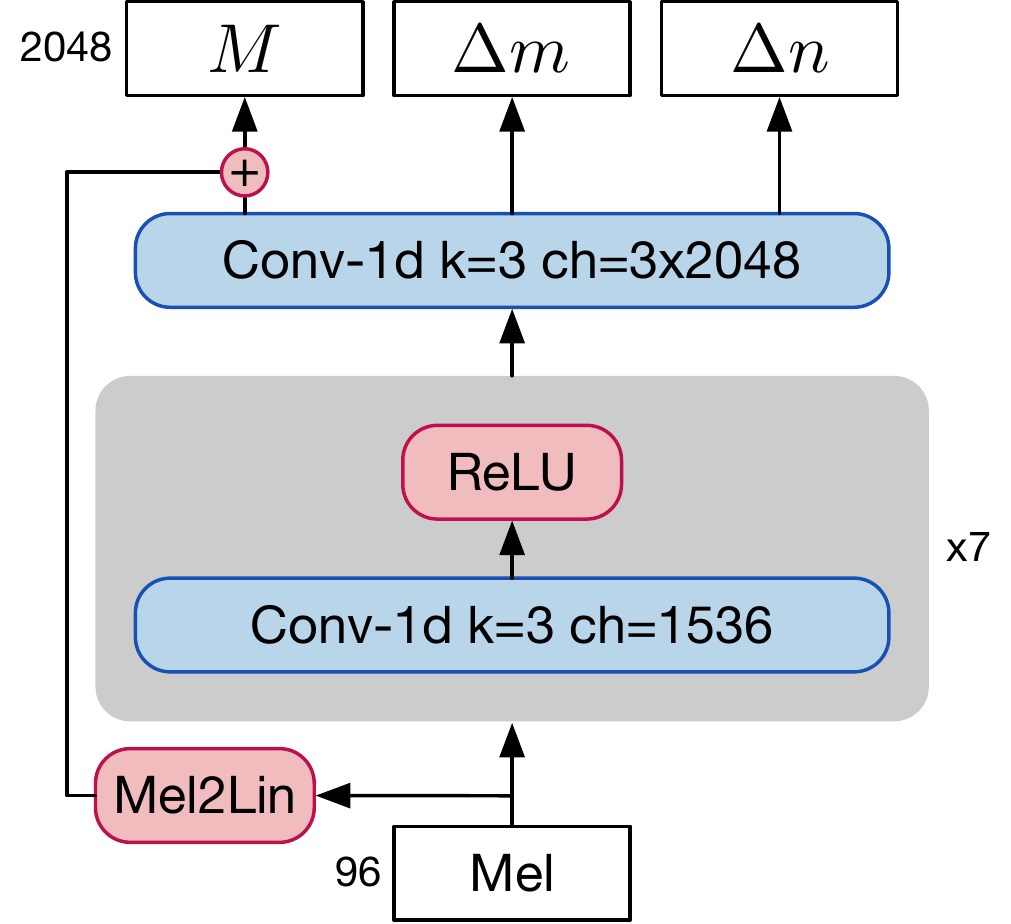}
  % \vspace{1em}
  \caption{
    Convolutional network architecture
    }
  \label{fig:network}
\end{figure}

The neural network is a stack of 8 1-d (time) convolutional layers with 
1536 hidden channels, a kernel size of 3 frames and ReLU activations (Fig. \ref{fig:network}).
% Even if the input and output tensors are 2d time-frequency representations,
% they are not frequency shift-invariant, therefore the convolution is only applied along the time dimension.
The input and output have the same number of time frames, but different numbers of 
frequency bins, and their center frequencies are different, i.e.\ log-spaced for the 
mel spectrogram input and linearly spaced for the magnitude and phase gradient outputs.
The frequency bins of the input and the magnitude channel of the output are independently standardized
using the mean and standard deviation values computed from the training set. We found that a 
direct path from
the input to the magnitude channel of the output leads to significant improvements in the 
reconstruction of magnitude and the training speed. This direct path consists only of a frequency
warping operation from mel- to linear-scale.

The phase gradients $(\phi'_i, \phi'_m)$ are computed using the 
Auger-Flandrin technique \cite{auger1995improving}.
Although we have not experimented with other ways to compute the phase gradient, it was argued 
\cite{fulop2006algorithms} that using a less precise method, such as finite 
differences might perform just as well.
Instead of using the phase gradients directly, we use the vertical and horizontal bin offsets
(Eq.\ \eqref{eq:n-bin-offset-deltas}), 
which are derived from the two components of the 
phase gradient. In order to remove outliers, the absolute bin offsets are clipped to $4.0$ on the frequency dimension and to 
$N/2R$ on the time dimension.

The model uses linear output activation on all output channels except for the magnitude channel,
where we apply a scaled $\tanh$ activation $f_{\beta}(x)=\beta\tanh(x / \beta)$ with 
$\beta=5$ to use mostly the linear regime while also preventing overflow \cite{engel2019gansynth}. 
The three channels are then scaled 
and offset appropriately to match the statistics of the targets. 
% The magnitude is first 
% inverse-standardized using the mean and standard deviation extracted from the magnitude values in dB
% from the training set, then converted from dB to linear amplitude. The phase gradient channels are 
% simply scaled by the values used for clipping the targets.

\subsubsection{Losses} \label{sssec:losses}

The magnitude channel is trained with a Mean Square Error (MSE) loss term,
and a further MSE loss term computed on the first 20 linear-frequency cepstral coefficients (LFCC).
\begin{eqnarray}
  \mathcal{L}_1 &=& (\hat{M} - M)^2  \\
  \mathcal{L}_2 &=& (\text{DCT}_{:20}(\hat{M}) - \text{DCT}_{:20}(M))^2,
\end{eqnarray}
where $\hat{M}$ indicates the estimated magnitude spectrum, $M$ the target magnitude spectrum, 
and $\text{DCT}$ the normalized Discrete Cosine Transform; 
in these and the following loss formulas the $[m, n]$ indices and the global average 
operation have been omitted for simplicity. While $\mathcal{L}_1$ acts on point estimates,
$\mathcal{L}_2$ pushes the spectral envelope towards its true value, 
leading to faster convergence and better reconstruction quality.
% LFCC: hs28c7ifr7
% no LFCC: tu78arzpwq

The phase gradient channels are trained with another MSE loss, weighted by the power spectrum
of the target signal $M^2$.
A matrix $\lambda \in [0, 1]$, computed from the phase gradient (Sect. \ref{ssec:methodphase}), 
is used to distinguish sinusoidal and impulsive components. 
The idea is that the phase derivative along the time/frequency dimension contributes to the loss only for
sinusoidal/impulsive components:
\begin{equation}
  \mathcal{L}_3 = 
  \begin{cases} 
    M^2 (\hat{\Delta{m}} - \Delta{m})^2 & \lambda > 0.5 \\
    M^2 (\hat{\Delta{n}} - \Delta{n})^2 & \lambda \le 0.5, \\
  \end{cases},
\end{equation}
where $(\hat{\Delta{m}}, \hat{\Delta{n}})$ are the estimated bin offsets.

Finally, because $\lambda$ is a function of $\nabla{\phi}$ and is used for 
integration, we add a loss term:
\begin{equation}
  \mathcal{L}_4 = M^2 (\hat{\lambda} - \lambda)^2,
\end{equation}
where $\hat{\lambda}$ is computed using the phase gradient estimates and 
$\lambda$ using the target values.
The final loss is a weighted sum of all the loss terms:
\begin{equation}
  \mathcal{L} = \sum_l \alpha_l \mathcal{L}_l,
\end{equation}
where all weights are set to 1 except $\alpha_2 = 0.1$ 
to balance the contribution of all terms during training.
% comparable/homogeneous/consistent

Differently from time-domain methods for mel spectrogram inversion, 
we found reconstruction losses to yield satisfying results and 
did not add any adversarial loss.
% given review comment to motivate more
Evaluating the advantages of including adversarial losses 
is left for future research.

\subsection{Phase integration}\label{ssec:methodphase}

The algorithm we use for integrating phase from phase gradients relies on the 
classification of spectral bins into either sinusoidal, transient or noise components.

The classification uses the phase gradient and relies on the following
rationale \cite{fitz2009unified}: around sinusoidal/impulsive components the reassigned frequency/time
is approximately constant along frequency/time

\begin{equation}
  \lambda[m, n] = e^{-(\frac{d}{dm}\dot{m}[m, n] / \frac{d}{dn}\dot{n}[m, n]) ^2},
\end{equation}
where the derivatives $d/dm$ and $d/dn$ are computed with centered finite differences.

After computing $\lambda$, the phase gradients are propagated horizontally 
(Eq.\ \eqref{eq:phigrad-if}) if $\lambda > \lambda^\text{S}$, 
vertically (Eq.\ \eqref{eq:phigrad-lgd}) if $\lambda < \lambda^\text{I}$, and set 
to a random value otherwise. $\lambda^\text{I}$ and $\lambda^\text{S}$ are 
threshold values for 
impulsive and sinusoidal components, and are used to identify the spectral bins
over which respectively vertical or horizontal phase coherence should be enforced.
We empirically set $\lambda^\text{I}=0.4$ and $\lambda^\text{S}=0.5$, 
as these values performed well on early trials.

% TODO:
% A comparison with other existing phase integration algorithms is presented in 
% Sect. \ref{sec:experiments}.

% TODO: paragraph on the comparison with other phase integration algorithms.

% Efficient, magnitude agnostic
% and differentiable.

\section{Experiments} \label{sec:experiments}

In this section we discuss how we evaluate the pitch stability of the proposed model,
comparing to strong baseline vocoder models from the speech synthesis literature.

\subsection{Experimental Setup} \label{ssec:experimental-setup}
We compare the reconstruction of the proposed {\tt phase-gradient} model against
state-of-the-art approaches: 
{\tt melgan} \cite{kumar2019melgan}\footnote{
  \href{https://github.com/descriptinc/melgan-neurips}{https://github.com/descriptinc/melgan-neurips}
  }, 
{\tt hifigan} \cite{kong2020hifi}\footnote{
  \href{https://github.com/kan-bayashi/ParallelWaveGAN.git}{https://github.com/kan-bayashi/ParallelWaveGAN.git}
  }, 
{\tt cargan} \cite{morrison2021chunked}\footnote{
  \href{https://github.com/descriptinc/cargan}{https://github.com/descriptinc/cargan}
  }, and 
{\tt diffwave} \cite{kong2020diffwave}\footnote{
  \href{https://github.com/lmnt-com/diffwave}{https://github.com/lmnt-com/diffwave}
  }.
All models have been trained on the same data, containing 13 hours of ambient music loops 
from commercial libraries\footnote{we used the following loop packs licensed from
Big Fish Audio Ltd.: Ambient Piano, Ambient Skyline 3, Ambient Waves, Eclipse: Ambient Guitars,
Ethereal Harp, Zen Ambient Vol. 2}, 
split into training, validation and test in the ratio of 80/10/10, which we refer to as the Ambient dataset.
The audio from these loop libraries is converted 
to mono at 44.1kHz, 16-bit, the spectrograms are computed with a 
frame size of 2048 samples and hop size of 256 samples, 
and finally 96 bands are used for the mel spectrogram. 
% We include a version of the {\tt phase-gradient} model using larger frames of 4096 samples, 
% named {\tt phase-gradient-4096}, to test the advantages of higher frequency resolution.

All neural networks were trained from scratch. 
The {\tt phase-gradient} network was trained using 2 Volta GPUs in parallel 
and batches of 32 examples, 
with the Adam optimizer and learning rate set to $3\mathrm{e}{-5}$. 
The training was stopped after 1024 epochs, when the 
validation loss converged, which took approximately 2 days.

% start training 2022-04-01:18:15:38
% epoch 1024 2022-04-03:18:14:35

As a further non machine-learned baseline, we consider a 
reconstruction algorithm {\tt griffin-lim},
which generates audio from the mel spectrogram by 
first warping the frequency axis and the values to obtain a magnitude spectrogram,
then applying the Griffin-Lim algorithm \cite{perraudin2013fast} for 500 iterations.

\begin{figure}[t]
  \centering
  \includegraphics[width=\columnwidth]{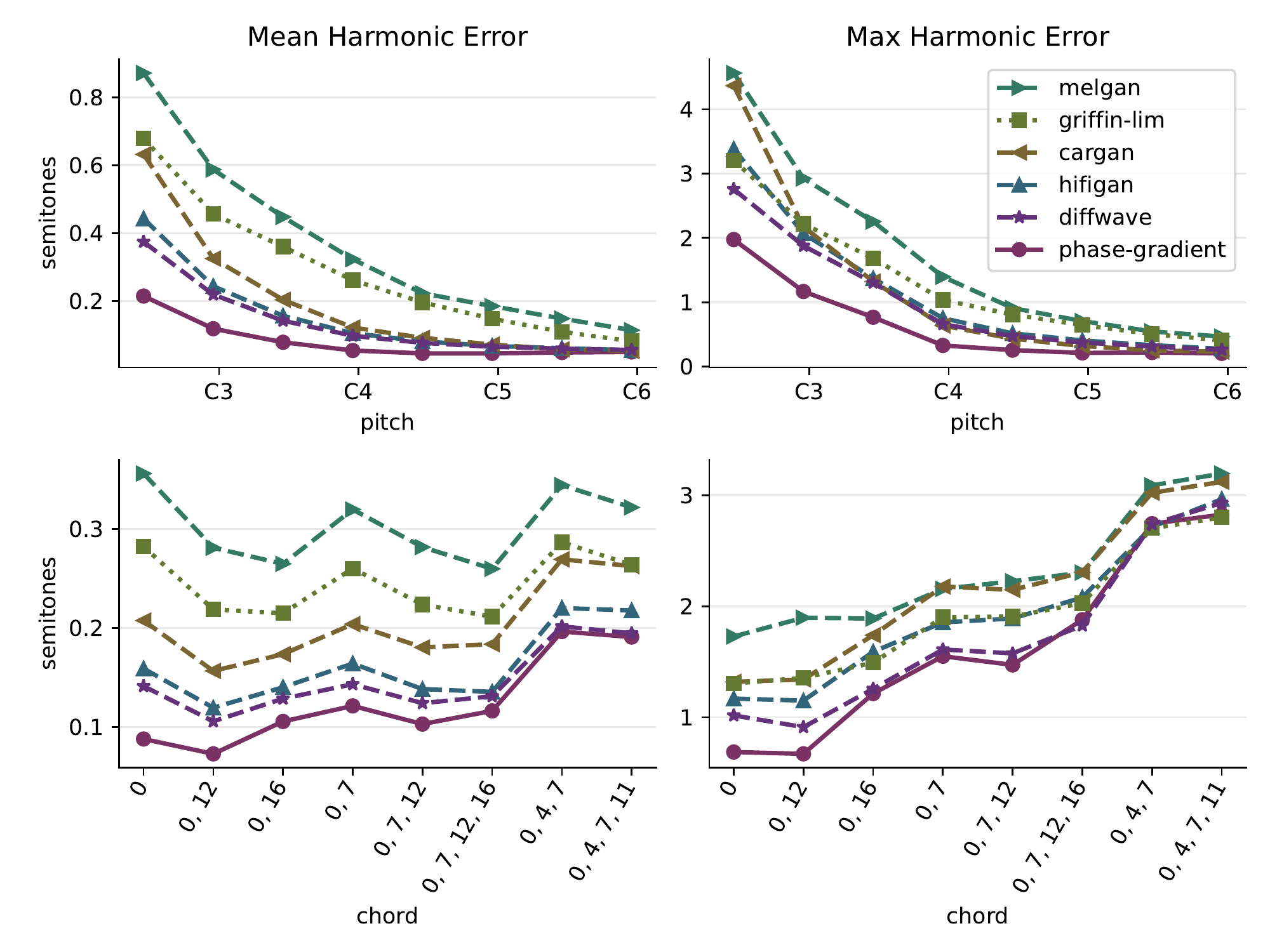}
  \vspace{-1em}
  \caption{
    Harmonic error of different 
    mel spectrogram inversion models 
    for synthesized notes in the (C2, C7) range (top row). 
    {\tt phase-gradient} model
    achieves lower error than the baseline models on the entire range. 
    The values have been smoothed with a moving average with size/stride 
    equal to 12/6 semitones to filter out noise.
    The bottom row shows the error when adding more notes in different combinations, 
    where the error is averaged over the entire pitch range, and 
    the numbers on the x axis indicate the intervals in semitones that
    are played simultaneously, e.g.\ "0, 4, 7" is the major triad.
    }
  \label{fig:pitch-error}
\end{figure}

\begin{figure}[t]
  \centering
  \includegraphics[width=\columnwidth]{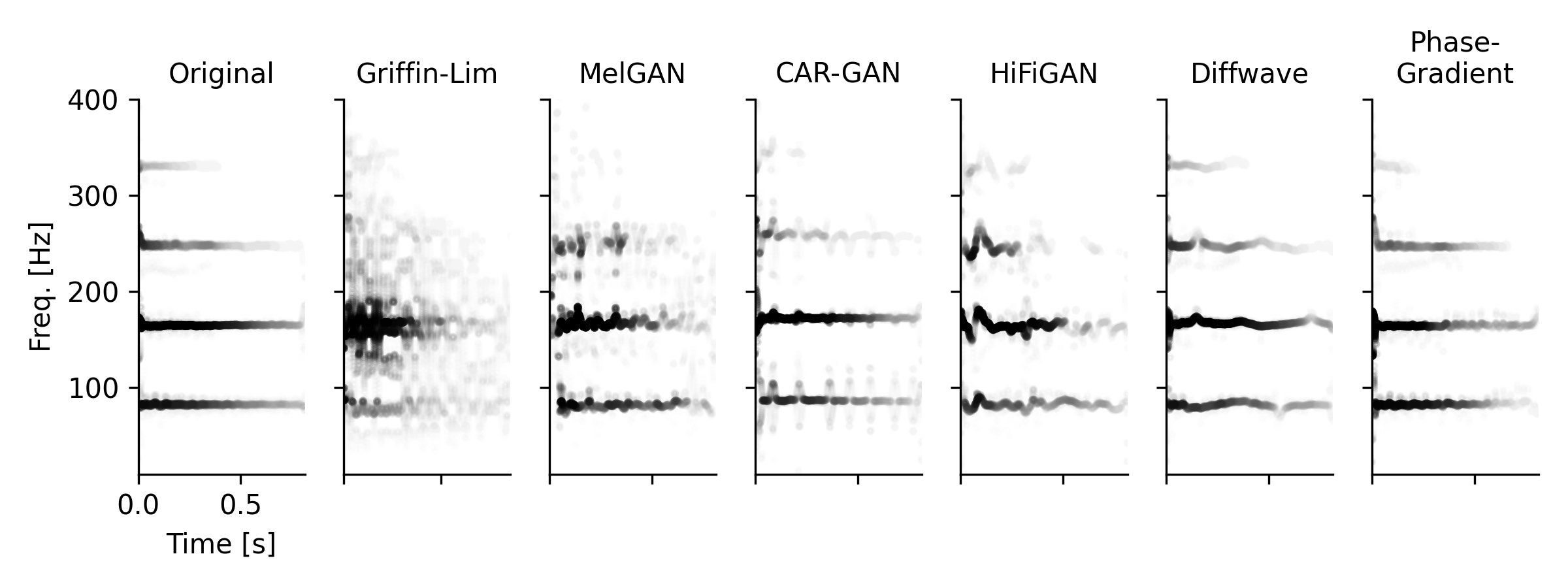}
  \vspace{-1em}
  \caption{
    Reconstructions of an E2 nylon guitar note using different mel spectrogram
    inversion models. The figure shows the reassigned spectrograms \cite{fitz2009unified} 
    which highlight the instability on the fundamental and the harmonics caused by the lack of 
    temporal phase coherence.
    }
  \label{fig:reassignedspecs}
\end{figure}

\subsection{Pitch stability} \label{ssec:results-pitch}

To evaluate the pitch stability of our model, 
we used FluidSynth\footnote{\href{http://www.fluidsynth.org}{http://www.fluidsynth.org}} to synthesize a dataset of
one second long notes and chords in the (C2, C7) range, 
using a set of four different sounds: a rhodes piano, a church organ, 
a string ensemble and a nylon guitar.

Measuring stability with a pitch tracker is only 
feasible for single notes, and we found that errors computed with a pitch tracker even for monophonic signals did not reflect our 
perception of reconstruction quality.
A possible explanation is that pitch-trackers average out errors in the 
harmonic frequencies, which is inconvenient in this context because 
the precise location of the overtones is important 
for timbre perception \cite{fletcher1962quality}.
% We argue that the stability of the harmonic frequencies is as important
% as the stability of the fundamental frequency. The ratios between the frequency of the 
% harmonics and the fundamental define the timbre \cite{fletcher1962quality} and are
% important to reconstruct coherent musical audio.

% define the harmonic error metric
For this reason, we define a harmonic error metric $H_\text{err}$ as the sum of the frequency errors
of the fundamental and the first 4 harmonic frequencies, expressed in the pitch scale:
\begin{equation}
  H_\text{err}[p, h, n] = 12 |\log_2(\frac{\hat{f}_{p,h}[n]}{f_{p,h}[n]})|,
\end{equation}
where $f_{p,h}[n]$ and $\hat{f}_{p,h}[n]$ are the frequencies of the $h$-th harmonic 
of the $p$-th note, at frame $n$,
for the original and the reconstructed audio respectively; the frequencies are estimated as
the closest peaks to the nominal frequency of the partial, using
quadratic interpolation, on a magnitude spectrum computed using 
frame/hop size equal to 4096/256 samples. We look at the mean and maximum values of the 
harmonic error, as a way to summarize the expected and worst case reconstruction
errors.

\begin{table*}[t] 
  \centering
  \begin{tabular}{l|rrr|rr|r|rr}
    \toprule
     & \multicolumn{3}{c|}{FAD ($\downarrow$)} & \multicolumn{2}{c|}{$H_\text{err}$ ($\downarrow$)} & \multicolumn{1}{c|}{MOS ($\uparrow$)} & \multicolumn{1}{c}{RTF ($\uparrow$)} & \multicolumn{1}{c}{\#Params} \\
     & Ambient & NSynth & N+C & Notes & Chords & \multicolumn{1}{c|}{NSynth} & & \\
    \midrule
    {\tt griffin-lim}\cite{perraudin2013fast}     &   10.59         &   6.16          &          6.79 &  0.28          &   0.24          & 1.42 $\pm$ 0.11 &   7.14 &  0 \\
    {\tt melgan}\cite{kumar2019melgan}            &    2.07         &   2.80          &          2.84 &  0.36          &   0.30          & 1.58 $\pm$ 0.15 & 179.90 &  4M \\
    {\tt cargan}\cite{morrison2021chunked}        &    6.47         &   8.31          &          8.45 &  0.21          &   0.21          & 1.74 $\pm$ 0.12 &   4.36 &  25M \\
    {\tt hifigan}\cite{kong2020hifi}              &  \textbf{0.85}  &   1.31          & \textbf{1.81} &  0.16          &   0.17          & 2.89 $\pm$ 0.12 &  68.12 &  13M \\
    {\tt diffwave}\cite{kong2020diffwave}         &    2.62         &   6.84          &          1.89 &  0.14          &   0.15          & 3.19 $\pm$ 0.12 &   0.32 &  7M \\
    {\tt phase-gradient}                          &    1.26         &   \textbf{1.26} &          1.86 &  \textbf{0.09} &   \textbf{0.14} & \textbf{3.73} $\pm$ 0.13 &  3.58 &  28M   \\
    \hline
    {\tt oracle}                                  &    0.51         &   0.33          &          0.00 &  0.00          &   0.00          & 4.34 $\pm$ 0.15 &    --- &  --- \\
    \bottomrule
  \end{tabular}
  \caption{Reconstruction results}
  \label{tab:results}
\end{table*}

Results show that our {\tt phase-gradient} model was able to reconstruct the 
single notes with lower mean and maximum harmonic error over the entire 
range of notes, especially in the lower pitch range (Fig. \ref{fig:pitch-error}(a)).
A possible explanation for the larger improvement on the low register is that
the long waveform period of lower-pitched notes makes them challenging to learn
in the time-domain, given the high number of phase variations.
A visualization of the different models' reconstruction of the same E2 guitar note
is provided in Fig.\ \ref{fig:reassignedspecs}.

We also synthesized different combinations of notes, to test the reconstruction
quality on more challenging signals. The combinations include octaves, major 12ths, perfect fifths,
an open- and a close-position voicing of the major triad, and a close-position voicing of the major seventh chord.
As expected, the harmonic error increases when adding more notes (Fig. \ref{fig:pitch-error}(b)), 
as they are harder to recognize in the input mel spectrogram. 
The {\tt phase-gradient} model is able to yield lower mean and maximum harmonic error 
on all the considered combinations of notes. However, the decrease in the reconstruction
quality, particularly on close-position chord voicings, suggests possible connections with
the target's frequency resolution.

% optionally note that the 4096 version is not lowest error only on one combination 
% ([0, 7, 12, 16]).

% TODO: Figure of reassigned spectrum of a reconstructed note with 
% Phase Gradient vs HiFiGAN

\subsection{Reconstruction quality} \label{ssec:results-rec}

To evaluate the overall reconstruction quality of the proposed model,
we compute the Frechét Audio Distance (FAD) \cite{kilgour2019frechet}
on the Ambient dataset, the dataset of 1920 one second long notes and chords (``N+C'') used in Sect.\ \ref{ssec:results-pitch},
and the NSynth dataset \cite{engel2017neural}.
The results are shown in Table \ref{tab:results} alongside
aggregated mean harmonic error results from the experiment discussed in Sect.\ \ref{ssec:results-pitch}.

The FAD metric compares embedding statistics generated on two potentially different sets of audio signals, 
i.e. evaluation and reference set.
On the Ambient and NSynth datasets we compute the FAD between the reconstructed test split (evaluation) and the original training split (reference).
On N+C, the reconstructed and original signals from the \emph{entire} dataset are used as evaluation and reference sets.
An ideal model ({\tt oracle}) is added to provide reference values, useful when the evaluation 
and reference sets are different.

The aggregated mean harmonic error results are computed over two meaningful subsets of the N+C dataset:
a ``Notes'' dataset, representing all single notes, and a ``Chords'' dataset,
containing all note combinations with more than one pitch class (see Fig.\ \ref{fig:pitch-error} bottom row).

% listening test
We conducted a small listening test to evaluate a set of notes reconstructed with different models.
The set includes G2 and G3 notes randomly selected from NSynth, for each instrument family.
The 5-scale Mean Opinion Score (MOS) values and the 95\% confidence intervals are shown in Table \ref{tab:results}.

The results show that the {\tt phase-gradient} model is competitive with other state-of-the-art 
models, despite having simpler neural network and training procedure.
The fact that {\tt hifigan} model is able to score lower FAD than the proposed model 
on the Ambient and N+C datasets suggests it has higher reconstruction accuracy on 
different sonic characteristics.

Specifically, we noticed that {\tt hifigan} was able to reconstruct impulsive
components such as transients and percussive onsets with higher energy 
and often more accurately than the {\tt phase-gradient} model, 
while struggling with the stability of pitched notes and chords.
The {\tt diffwave} model exhibited high frequency ``hissing''
noise, but was otherwise surprisingly stable on harmonic components. This quality 
likely stems from the wide \textasciitilde7s receptive field, 
obtained using the entire reverse process at generation time instead of a fast sampling schedule \cite{kong2020diffwave}.
As reported by its authors in \cite{morrison2021chunked}, 
we confirm that the reconstructions made by the {\tt cargan} model contained 
``boundary artifacts that appear as repeated clicks'', compromising their usability.
Finally, the reconstructions obtained with the {\tt melgan} model were characterized by 
heavy phase artifacts such as metallic sounds, and very unstable pitch.

% comment on generation time
Generation time is also shown in Table \ref{tab:results} in terms of real-time factor (RTF),
defined as the number of seconds of audio that can be generated per second, evaluated on a single
NVidia Volta GPU. 
{\tt phase-gradient}'s generation is faster than real time, and close to {\tt cargan}.
While {\tt phase-gradient}'s neural network is as fast as {\tt hifigan} and {\tt melgan},
99\% of the generation time is spent during the auto-regressive 
phase integration stage, suggesting a clear direction for optimization.

% Best models phase-gradient, hifigan, diffwave
% hifigan: better transient reconstruction, higher energy, slightly wobbly pitch
% diffwave hissing noise high frequency, good pitch stability (particularly in NSynth) due to the 
%   very long receptive field (note that we're using the full noise schedule on generation too)
% cargan: pulsating noise clicks
% melgan: worse phase artifacts with metalling sounding voice and very unstable pitch.

% Start with what audio features make HiFiGAN better than phase-gradient
% continue with reconstruction characteristics for other models

% 2. insights on model performances, e.g. cargan & diffwave
%  (reconstruct and play from an interactive bolt task)

% between the original audio of the training set and the reconstructed audio of the
% test set. The results (Fig.\ \ref{fig:results-rec}) show that the proposed
% model is competitive with the state of the art, although scoring lower than
% {\tt hifigan}. The performance gap could be attributed to the reconstruction
% of the transients and, more generally, the non-harmonic content of the ambient music loops.

% \begin{figure}[ht]
%   \centering
%   \includegraphics[width=\columnwidth]{figs/fad.pdf}
%   \vspace{-1em}
%   \caption{
%     Reconstruction quality comparison.
%     The dashed vertical line indicates the theoretical minimum, 
%     computed using the original audio from the test set.
%   }
%   \label{fig:results-rec}
% \end{figure}

\section{Conclusions} \label{sec:conclusions}

In this work we have proposed a new mel spectrogram inversion model designed for music
that achieves improved reconstruction of sustained notes and chords, compared to
state-of-the-art models from the speech synthesis literature.
This improvement is obtained using a frequency-domain 
target representation
that is time shift invariant for harmonic signal components.
% results
% We defined the harmonic error metric to evaluate the stability of reconstructed 
% notes and chords.
% The notes/chords reconstructed by the proposed model are consistently 
% more stable, while maintaining an overall 
% reconstruction quality competitive with the .
% than all the existing models
The proposed model is able to reconstruct single notes and chords 
respectively 60\% and 10\% more precisely than existing models, when evaluated with a
novel harmonic error metric, while still being competitive on 
generic loop reconstruction.
% overall reconstruction
Potential directions for improvement include using 
pitch-shift augmentation, investigating log-frequency target representations,
and training a separate time-domain model to supply the percussive components.

% The reconstruction of musical loops is competitive with the state of the art,
% with differences pointing to potential areas of further improvement.
% Assuming that time-domain models are better 
% such as the possibility of using the proposed model as the harmonic component of a composite model,
% where the percussive component is provided by a classic time-domain model.

% Improve generalisation with pitch-shift augmentation
% Use a logarithmic-frequency output space for the
% Time-Shift-invariance holds for time-derivative but not for freq-derivative
% Test the benefit of learning the partial derivative along the frequency axis
% Loss of presence due to potential phase cancellation of spectral bins 
% contributing to the same sinusoidal component 

\section{Acknowledgements}\label{sec:acknowledgements}
We would like to thank the following colleagues for their 
valuable help and feedback: David Varas Gonzalez, Tim O'Brien, 
Avery Wang, Meghna Ranjit, and André Bergner.

% For bibtex users:
\bibliography{ISMIRtemplate}

\begin{thebibliography}{10}
\providecommand{\url}[1]{#1}
\csname url@samestyle\endcsname
\providecommand{\newblock}{\relax}
\providecommand{\bibinfo}[2]{#2}
\providecommand{\BIBentrySTDinterwordspacing}{\spaceskip=0pt\relax}
\providecommand{\BIBentryALTinterwordstretchfactor}{4}
\providecommand{\BIBentryALTinterwordspacing}{\spaceskip=\fontdimen2\font plus
\BIBentryALTinterwordstretchfactor\fontdimen3\font minus
  \fontdimen4\font\relax}
\providecommand{\BIBforeignlanguage}[2]{{%
\expandafter\ifx\csname l@#1\endcsname\relax
\typeout{** WARNING: IEEEtran.bst: No hyphenation pattern has been}%
\typeout{** loaded for the language `#1'. Using the pattern for}%
\typeout{** the default language instead.}%
\else
\language=\csname l@#1\endcsname
\fi
#2}}
\providecommand{\BIBdecl}{\relax}
\BIBdecl

\bibitem{kumar2019melgan}
K.~Kumar, R.~Kumar, T.~de~Boissiere, L.~Gestin, W.~Z. Teoh, J.~Sotelo,
  A.~de~Br{\'e}bisson, Y.~Bengio, and A.~C. Courville, ``Melgan: Generative
  adversarial networks for conditional waveform synthesis,'' \emph{Advances in
  neural information processing systems}, vol.~32, 2019.

\bibitem{kong2020hifi}
J.~Kong, J.~Kim, and J.~Bae, ``Hifi-gan: Generative adversarial networks for
  efficient and high fidelity speech synthesis,'' \emph{Advances in Neural
  Information Processing Systems}, vol.~33, pp. 17\,022--17\,033, 2020.

\bibitem{oord2016wavenet}
A.~v.~d. Oord, S.~Dieleman, H.~Zen, K.~Simonyan, O.~Vinyals, A.~Graves,
  N.~Kalchbrenner, A.~Senior, and K.~Kavukcuoglu, ``Wavenet: A generative model
  for raw audio,'' in \emph{ISCA Speech Synthesis Workshop (SSW)}, 2016.

\bibitem{kalchbrenner2018wavernn}
N.~Kalchbrenner, E.~Elsen, K.~Simonyan, S.~Noury, N.~Casagrande, E.~Lockhart,
  F.~Stimberg, A.~v.~d. Oord, S.~Dieleman, and K.~Kavukcuoglu, ``Efficient
  neural audio synthesis,'' in \emph{International Conference on Machine
  Learning (ICML)}, 2018.

\bibitem{morrison2021chunked}
M.~Morrison, R.~Kumar, K.~Kumar, P.~Seetharaman, A.~Courville, and Y.~Bengio,
  ``Chunked autoregressive gan for conditional waveform synthesis,'' in
  \emph{International Conference on Learning Representations (ICLR)}, 2022.

\bibitem{engel2019gansynth}
J.~Engel, K.~K. Agrawal, S.~Chen, I.~Gulrajani, C.~Donahue, and A.~Roberts,
  ``Gansynth: Adversarial neural audio synthesis,'' in \emph{International
  Conference on Learning Representations (ICLR)}, 2019.

\bibitem{fitz2009unified}
K.~R. Fitz and S.~A. Fulop, ``A unified theory of time-frequency
  reassignment,'' \emph{arXiv preprint arXiv:0903.3080}, 2009.

\bibitem{kong2020diffwave}
Z.~Kong, W.~Ping, J.~Huang, K.~Zhao, and B.~Catanzaro, ``Diffwave: A versatile
  diffusion model for audio synthesis,'' in \emph{International Conference on
  Learning Representations (ICLR)}, 2021.

\bibitem{kandpal2022music}
N.~Kandpal, O.~Nieto, and Z.~Jin, ``Music enhancement via image translation and
  vocoding,'' in \emph{IEEE International Conference on Acoustics, Speech and
  Signal Processing (ICASSP)}, 2022, pp. 3124--3128.

\bibitem{pruvsa2016real}
Z.~Pru{\v{s}}a and P.~L. S{\o}ndergaard, ``Real-time spectrogram inversion
  using phase gradient heap integration,'' in \emph{International Conference on
  Digital Audio Effects (DAFx)}, 2016, pp. 17--21.

\bibitem{dolson1986phase}
M.~Dolson, ``The phase vocoder: A tutorial,'' \emph{Computer Music Journal},
  vol.~10, no.~4, pp. 14--27, 1986.

\bibitem{pruuvsa2017phase}
Z.~Pru{\v{s}}a and N.~Holighaus, ``Phase vocoder done right,'' in \emph{IEEE
  European Signal Processing Conference (EUSIPCO)}, 2017, pp. 976--980.

\bibitem{laroche1999improved}
J.~Laroche and M.~Dolson, ``Improved phase vocoder time-scale modification of
  audio,'' \emph{IEEE Transactions on Speech and Audio Processing}, vol.~7,
  no.~3, pp. 323--332, 1999.

\bibitem{damskagg2017audio}
E.-P. Damsk{\"a}gg and V.~V{\"a}lim{\"a}ki, ``Audio time stretching using fuzzy
  classification of spectral bins,'' \emph{Applied Sciences}, vol.~7, no.~12,
  p. 1293, 2017.

\bibitem{auger1995improving}
F.~Auger and P.~Flandrin, ``Improving the readability of time-frequency and
  time-scale representations by the reassignment method,'' \emph{IEEE
  Transactions on signal processing}, vol.~43, no.~5, pp. 1068--1089, 1995.

\bibitem{fulop2006algorithms}
S.~A. Fulop and K.~Fitz, ``Algorithms for computing the time-corrected
  instantaneous frequency (reassigned) spectrogram, with applications,''
  \emph{Journal of the Acoustical Society of America}, vol. 119, no.~1, pp.
  360--371, 2006.

\bibitem{perraudin2013fast}
N.~Perraudin, P.~Balazs, and P.~L. S{\o}ndergaard, ``A fast griffin-lim
  algorithm,'' in \emph{IEEE Workshop on Applications of Signal Processing to
  Audio and Acoustics}, 2013, pp. 1--4.

\bibitem{fletcher1962quality}
H.~Fletcher, E.~D. Blackham, and R.~A. Stratton, ``Quality of piano tones,''
  \emph{Journal of the Acoustical Society of America}, vol.~34, pp. 749--761,
  1962.

\bibitem{kilgour2019frechet}
K.~Kilgour, M.~Zuluaga, D.~Roblek, and M.~Sharifi, ``Fr{\'e}chet audio
  distance: A reference-free metric for evaluating music enhancement
  algorithms,'' in \emph{ISCA Interspeech}, 2019, pp. 2350--2354.

\bibitem{engel2017neural}
J.~Engel, C.~Resnick, A.~Roberts, S.~Dieleman, M.~Norouzi, D.~Eck, and
  K.~Simonyan, ``Neural audio synthesis of musical notes with wavenet
  autoencoders,'' in \emph{International Conference on Machine Learning
  (ICML)}, 2017.

\end{thebibliography}

% For non bibtex users:
%\begin{thebibliography}{citations}
% \bibitem{Author:17}
% E.~Author and B.~Authour, ``The title of the conference paper,'' in {\em Proc.
% of the Int. Society for Music Information Retrieval Conf.}, (Suzhou, China),
% pp.~111--117, 2017.
%
% \bibitem{Someone:10}
% A.~Someone, B.~Someone, and C.~Someone, ``The title of the journal paper,''
%  {\em Journal of New Music Research}, vol.~A, pp.~111--222, September 2010.
%
% \bibitem{Person:20}
% O.~Person, {\em Title of the Book}.
% \newblock Montr\'{e}al, Canada: McGill-Queen's University Press, 2021.
%
% \bibitem{Person:09}
% F.~Person and S.~Person, ``Title of a chapter this book,'' in {\em A Book
% Containing Delightful Chapters} (A.~G. Editor, ed.), pp.~58--102, Tokyo,
% Japan: The Publisher, 2009.
%
%
%\end{thebibliography}

\end{document}